\documentclass[10pt,journal,a4paper]{IEEEtran}

\usepackage{amssymb}
\usepackage{amsmath}
\usepackage{graphicx}
\usepackage{subfigure}
\usepackage{cite}
\usepackage{enumerate}
\usepackage{bm}
\usepackage{epstopdf}
\usepackage{multirow}

\newcommand{\beq}{\begin{equation}}
\newcommand{\eeq}{\end{equation}}

\begin{document}

\title{Full-duplex MAC Protocol Design and Analysis}

\author{
\IEEEauthorblockN{\small{Yun Liao}\IEEEauthorrefmark{1}, \small{Kaigui Bian}\IEEEauthorrefmark{1}, \small{Lingyang Song}\IEEEauthorrefmark{1}, \small{and Zhu Han}\IEEEauthorrefmark{2} \\}
\IEEEauthorblockA{\IEEEauthorrefmark{1}\small{School of Electrical Engineering and Computer Science, Peking University, Beijing, China,}\\
\IEEEauthorrefmark{2}\small{Electrical and Computer Engineering Department, University of Houston, Houston, TX, USA.}\\}
}

%\author{Author 1, Author 2, Author 3, and Author 4}

%-make the title area
\maketitle

\thispagestyle{empty}
\pagestyle{empty}

\begin{abstract}
The idea of in-band full-duplex~(FD) communications revives in recent years owing to the significant progress in the self-interference cancellation and hardware design techniques, offering the potential to double spectral efficiency. The adaptations in upper layers are highly demanded in the design of FD communication systems. In this letter, we propose a novel medium access control~(MAC) using FD techniques that allows transmitters to monitor the channel usage while transmitting, and backoff as soon as collision happens. Analytical saturation throughput of the FD-MAC protocol is derived with the consideration of imperfect sensing brought by residual self-interference~(RSI) in the PHY layer. Both analytical and simulation results indicate that the normalized saturation throughput of the proposed FD-MAC can significantly outperforms conventional CSMA/CA under various network conditions.

\end{abstract}

\IEEEpeerreviewmaketitle

\vspace{-1em}
%%%%%%%%%%%%%%%%%%%%%%%
\section{Introduction}%
%%%%%%%%%%%%%%%%%%%%%%%

The idea of single-channel full-duplex~(FD) communication, where a node can transmit and receive using the same time and frequency resources has been proposed for over a decade because of its potential of double capacity. These several years witness the revival of FD research thanks to the development of self-interference cancellation and suppression techniques~\cite{jain2011practical}. In addition to the increasing interest and rapid development in the PHY layer realization, it also requires adaptations in protocol design for the upper
layers like the medium access control~(MAC) layer.

Some full-duplex MAC protocols have been proposed during the last few years \cite{jain2011practical,sahai2011pushing,singh2011efficient,goyal2013distributed}. In \cite{jain2011practical} and \cite{sahai2011pushing}, the centralized FD-MAC protocols are proposed: \cite{jain2011practical} considers bidirectional transmission between a pair of nodes, and uses busytone to eliminate the hidden terminal problem, and \cite{sahai2011pushing} design the protocol with three new elements, namely, shared random backoff, header snooping and virtual backoffs. Decentralized FD-MAC protocols are proposed in \cite{singh2011efficient,goyal2013distributed} based on CSMA/CA.
The former mainly focuses on bidirectional transmission between users, and the latter discusses simultaneous transmissions among two or three FD users. However, the way how to fully utilize FD techniques for distributed wireless networks with multiple contending users, and comprehensive analysis from PHY to MAC layers still requires further investigation.

To this end, in this letter, we propose a new distributed FD-MAC protocol for decentralized FD communication networks in which a number of FD transceiver pairs compete for transmission opportunities. Due to FD techniques, transmitters are able to sense and monitor the channel usage state while they are transmitting. Thus, if collision happens, users are able to backoff in short time before finishing the whole data packet and waiting for the absence of the ACK packet. We derive the analytical throughput of the proposed FD-MAC protocol by taking imperfect sensing caused by residual self-interference~(RSI) into consideration \cite{mypaper}. Both analytical and simulation evaluations show that the proposed protocol can significantly improve system throughput and achieve high normalized saturation throughput.

Note that different from the CSMA/CD in Ethernet, the proposed FD-MAC operates in wireless environment with fading channels and RSI caused by FD techniques. Thus, the sensing performance is imperfect and the design and analysis of the FD-MAC is non-trivial.

\vspace{-1em}
%%%%%%%%%%%%%%%%%%%%%%%%%%%%%%%%%%%%%%%%%%%%%%%
\section{FD-MAC Protocol}%
%%%%%%%%%%%%%%%%%%%%%%%%%%%%%%%%%%%%%%%%%%%%%%%

\begin{figure*}[!t]
\centering
\includegraphics[width=5.2in]{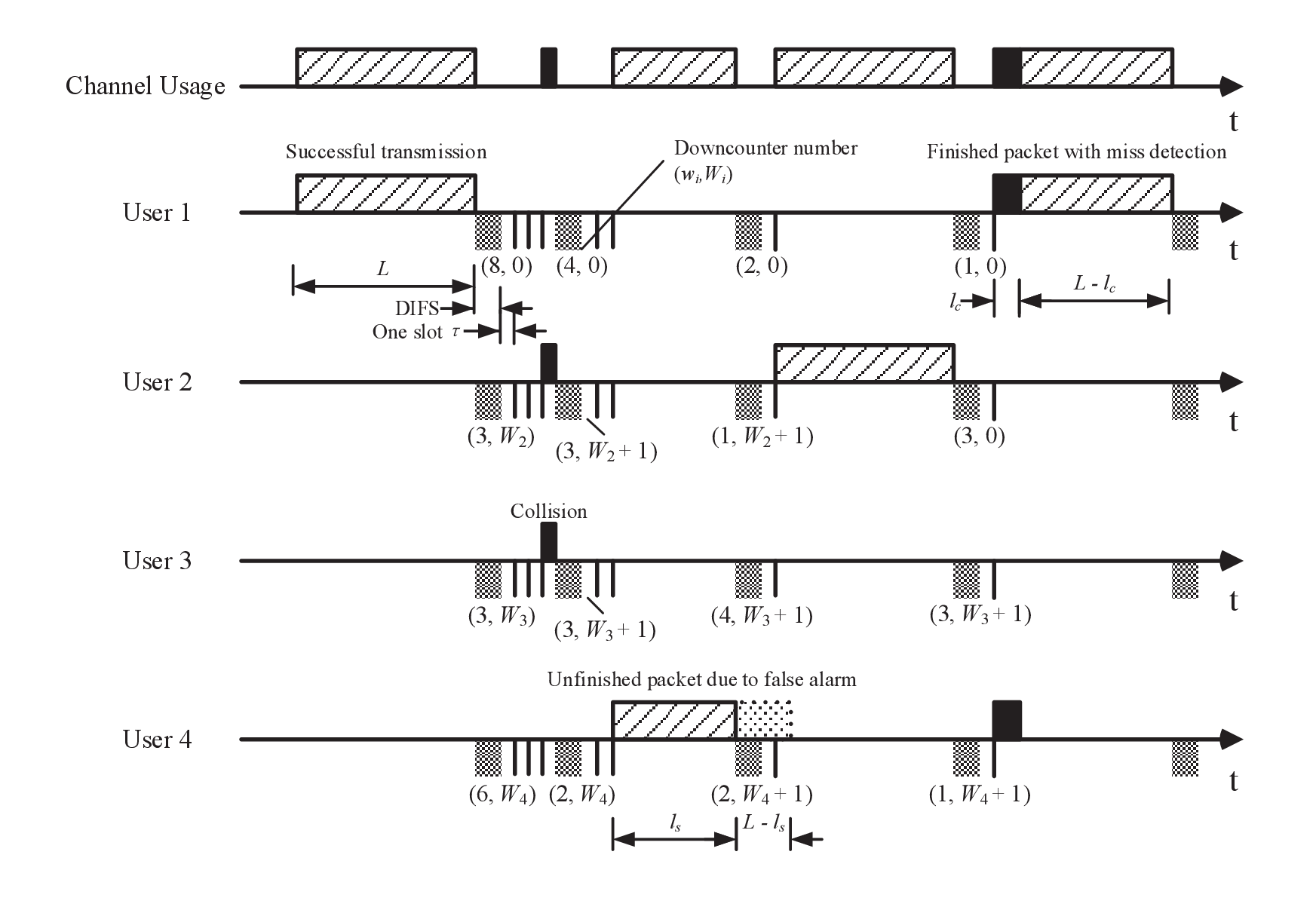}
\vspace{-2em}
\caption{FD-MAC Protocol in decentralized FD networks, in which $\left(w_i,W_i\right)$ denotes the residual backoff time and the backoff stage of user $i$.} \label{protocol}
%\vspace{-0.5em}
\end{figure*}

In this section, we elaborate the proposed FD-MAC protocol based on FD techniques and carrier sense multiple access with collision avoidance~(CSMA/CA) concepts, such that each user can sense the spectrum and determine whether other users are occupying it while transmitting its own data. This feature leads to the possibility of a new MAC protocol design in which users are not assumed ``blind'' when transmitting.

\vspace{-1em}
\subsection{System Model}
We consider a system with $M$ FD-enabled contending users operating under the saturated condition, i.e., each user always has a packet to transmit. All users sense the target channel continuously to determine whether other users are occupying it. FD techniques allow users to keep sensing the channel while transmitting. Thus, users can detect channel state change swiftly, and change their own states accordingly.

Note that the sensing in FD users can never be perfect due to the impact of RSI. In analysis of conventional CSMA, noise term is neglected, and sensing is commonly assumed perfect. For simplicity and comparison fairness, we also omit noise term in sensing, and only consider the impact of RSI \cite{mypaper}. Furthermore, we assume that when more than two users collide, the received signal from other users overwhelms the RSI, and the sensing is perfect. Thus, the effects of imperfect sensing including false alarm and miss detection\footnote{The values of false alarm and miss detection are related to the RSI and slot length in PHY~\cite{mypaper}. Generally, they both increase with the RSI, i.e., the sensing performance gets degraded with self-interference. Also, there exist requirements for minimum slot length to achieve reliable sensing, which are related to sampling frequency in sensing, channel conditions, number of users, and the RSI level, etc.} are considered only in the following two cases, respectively: (1) one single user occupies the channel; (2) two users collide with each other.

\vspace{-1em}
\subsection{Protocol Design}

Fig.~\ref{protocol} shows the proposed protocol, which consists of the following several parts.

\textbf{Sensing:} All users keep sensing the channel continuously regardless of its own activity, and make decisions of the channel usage at the end of each slot with duration $\tau$, which is the required time to reliably detect the transmission of any other user.

\textbf{Backoff mechanism:} Once the channel is judged idle without interruption for a certain period of time as long as a distributed interference space~(DIFS) (shown as the dotted area below each line), users check their own backoff timers and generate a random backoff time for additional deferral if their timers have counted down to zero. The additional backoff time after a DIFS is also slotted by $\tau$, i.e., the backoff time is expressed as
\beq
\text{Backoff Time}=w\times\tau=\text{Random}\left(\text{CW}\right)\times \tau,
\eeq
where CW$=2^{W}\cdot\text{CW}_{\min}$ is the contention window length, and $w=\text{Random}\left(\text{CW}\right)$ is a random integer drawn from the uniform distribution over the interval $\left[0,\text{CW}\right)$, where $W\in\left[0,W_{\max}\right]$ is the backoff stage depending on the number of unsuccessful transmissions for a packet. The countdown starts right after the DIFS, and suspends when the channel is detected occupied by others.

\textbf{Channel access and transmission suspension:} A user accesses the channel and begins transmission when its timer reaches zero. During the transmission, if it detects the signal from other users, it stops its transmission and switches to the backoff procedure immediately. If the packet is finished, the user resets the the backoff state $W = 0$. Otherwise, it sets $W=\min\left\{W+1, W_{\max}\right\}$.

\vspace{-1em}
%%%%%%%%%%%%%%%%%%%%%%%%%%%%%%%%%%%%%%%%%%%%%%%
\section{Performance Analysis}%
%%%%%%%%%%%%%%%%%%%%%%%%%%%%%%%%%%%%%%%%%%%%%%%

In this section, we study the analytical performance of the proposed FD-MAC protocol, and derive its saturation throughput. Note that when only one user is transmitting, all other users can detect its transmission perfectly, which means that once a collision-free transmission begins, it either completes the packet or suspends it because of false alarm. This process is independent with other users' sensing and contending, and thus, contention and transmission can be considered separately. We first derive the probability that one transmission attempt collides with other transmissions. Then by considering the average successful transmission length, we evaluate the saturation throughput.

\vspace{-1em}
\subsection{Collision Probability}

\begin{figure}[!t]
\center
\includegraphics[width=3in]{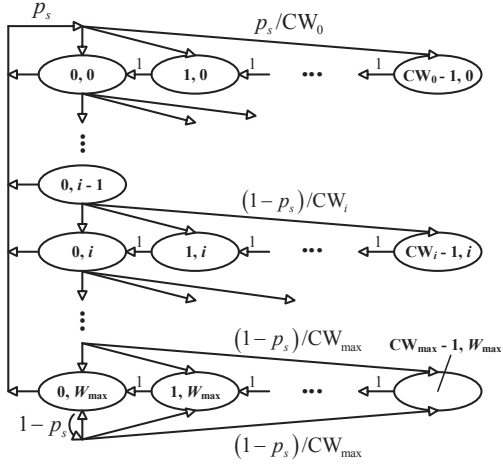}
\vspace{-1.5em}
\caption{Markov chain of the backoff window size.} \label{Markov}
\vspace{-2.5em}
\end{figure}

We follow the assumption in \cite{bianchi2000performance} that each packet gets collided with a same probability independent of the value of CW$_i$. Let $\{w_i,W_i\}$ denote the state of the $i^{th}$ contending user. For each user, the state change can be modeled as a discrete-time Markov chain illustrated in Fig.~\ref{Markov}. The non-zero transition probabilities are given as
\beq
\left\{ {\begin{array}{*{20}{l}}
&P\left(w_i-1,W_i|w_i,W_i\right)=1,\\
&~~~~~~~~~~~~~~~~w_i \in \left(0,\text{CW}_i\right),W_i = \left[0,W_{\max}\right],\\
&P\left(w_i,0|0,W_i\right)=p_s/\text{CW}_{\min},\\
&~~~~~~~~~~~~~~~~w_i \in \left[0,\text{CW}_{\min}\right),W_i = \left[0,W_{\max}\right],\\
&P\left(w_i,W_i+1|0,W_i\right)=\left(1-p_s\right)/\text{CW}_{i+1},\\
&~~~~~~~~~~~~~~~~w_i \in \left[0,\text{CW}_{i+1}\right),W_i = \left[0,W_{\max}\right),\\
&P\left(w_i,W_{\max}|0,W_{\max}\right)=\left(1-p_s\right)/\text{CW}_{\max},\\
&~~~~~~~~~~~~~~~~w_i \in \left[0,\text{CW}_{\max}\right),\\
\end{array}} \right.
\eeq
where $p_s$ denotes the probability that the considered user successfully finishes its transmission without awareness of collision. Note that $p_s$ does not equal to the non-collision probability due to imperfect sensing. Specifically, if two users collide, it is possible that only one user stops, and when one user is transmitting without collision, it may cease the transmission due to false alarm.

Consider the steady-state distribution of the Markov chain, the probability that one user stays in each state can be calculated. Let $p_{w,W}$ denote the probability that one user is in the state of $\{w,\text{W}\}$, and the probability that a certain user begins transmission in the next slot is

\begin{small}
\begin{equation}\label{p}
\begin{split}
&p = \sum\limits_{W = 0}^{W = {W_{\max }}} {{p_{0,W}}} \\
 &= \frac{{2\left( {2{p_s} - 1} \right)}}{{\left( {2{p_s} - 1} \right)\left( {{\rm{C}}{{\rm{W}}_{\min }} + 1} \right) + \left( {1 - {p_s}} \right){\rm{C}}{{\rm{W}}_{\min }}\left( {1 - {{\left( {2 - 2{p_s}} \right)}^{{W_{\max }}}}} \right)}}.
\end{split}
\end{equation}
\end{small}

Then, consider the relation between $p_s$ and $p$. For simplicity, we assume the packet length $L$ is fixed. The calculation of $p_s$ has two pre-requisites:
\begin{enumerate}
\item The probability that one user starts collision-free transmission after colliding with others for $l$ slots, denoted as $p_a\left(l\right),l\in\left[0,L\right]$, which can be expressed as
    \beq
    {p_a}\left( l \right) = \left\{ {\begin{array}{*{20}{l r}}
    {{{\left( {1 - p} \right)}^{M - 1}}}&{l = 0,}\\
    \multicolumn{2}{l}{\left( {M - 1} \right)p{{\left( {1 - p} \right)}^{M - 2}}P_m^{2l - 1}\left( {1 - {P_m}} \right)~~~~~}\\
    \multicolumn{2}{r}{1 \le l \le L-1,}\\
    {\left( {M - 1} \right)p{{\left( {1 - p} \right)}^{M - 2}}P_m^{2L - 1}}&{l = L.}
    \end{array}} \right.
    \eeq
\item The probability of successfully finishing current packet with residual collision-free length of $l$, denoted as $p_b\left(l\right),l\in\left[0,L\right]$:
    \beq
    {p_b}\left( l \right) = \begin{array}{*{20}{c}}
    {{{\left( {1 - {P_f}} \right)}^l}}&{0 \le l \le L}.
    \end{array}
    \eeq
\end{enumerate}

Successful transmission requires at least one user transmits the entire packet without the awareness of collision. Thus, $p_s$ can be calculated as
\beq\label{ps}
\begin{split}
{p_s} &= \sum\limits_{l = 0}^{L} {{p_a}\left( l \right){p_b}\left( {L - l} \right)}\\
&={\left( {1 - p} \right)^{M - 1}}{\left( {1 - {P_f}} \right)^L} \\
 &+ \left(M - 1\right)p{\left( {1 - p} \right)^{M - 2}}P_m\cdot\frac{\left(1 - P_f\right)^L - P_m^{2L}}{1 - P_f - P_m^2}.
\end{split}
\eeq

Combining \eqref{p} and \eqref{ps}, the values of $p$ and $p_s$ can be solved numerically.

\vspace{-1em}
\subsection{Throughput}
We use the time fraction that the channel is occupied for successful transmission as the normalized throughput, i.e., the throughput is defined as,
\beq\label{c}
\begin{split}
C &= \frac{\mathbb{E}\left[\text{Successful transmission length}\right]}{\mathbb{E}\left[\text{Consumed time for a successful transmission}\right]}\\
&= \frac{P_sL_s}{P_e+P_s\left(L_s+\text{DIFS}\right)+P_c\left(L_c+\text{DIFS}\right)},
\end{split}
\eeq
where $P_s = Mp\left(1-p\right)^{M-1}$ denotes the probability that a successful transmission occurs, $P_e = \left(1-p\right)^M$ is the probability that the channel is empty, $P_c = 1 - P_e - P_s$ represents the collision probability, and $L_s,~L_e$, and $L_c$ denote the average length of successful transmission, empty state, and collision, respectively. The average length of successful transmission and collision can be calculated as, respectively,
\beq\label{ls}
\begin{split}
{L_s} &= \sum\limits_{l = 1}^{L - 1} {l{{\left( {1 - {P_f}} \right)}^{l - 1}}{P_f}}  + L{\left( {1 - {P_f}} \right)^{L - 1}} \\
&= \frac{{1 - {{\left( {1 - {P_f}} \right)}^{L - 1}}}}{{{P_f}}} + {\left( {1 - {P_f}} \right)^{L - 1}},
\end{split}
\eeq
\beq\label{lc}
\begin{split}
{L_c}& = \left( {{P_c} + \left( {\begin{array}{*{20}{c}}
M\\
2
\end{array}} \right){p^2}{{\left( {1 - p} \right)}^{M - 2}}\sum\limits_{l = 1}^{L - 1} {P_m^{2l}\left( {1 - P_m^2} \right)l} } \right)/{P_c}\\
& = 1 + \left( {\begin{array}{*{20}{c}}
M\\
2
\end{array}} \right){p^2}{\left( {1 - p} \right)^{M - 2}}\frac{{P_m^2\left( {1 - P_m^{2L - 2}} \right)}}{{{P_c}\left( {1 - P_m^2} \right)}}.
\end{split}
\eeq
The throughput is readily obtained by substituting \eqref{ls} and \eqref{lc} into \eqref{c}.

\vspace{-1em}
\subsection{Comparison with the Basic CSMA/CA Mechanism}

We make a comparison between the proposed FD-MAC protocol with the conventional CSMA/CA in this subsection. For fairness, we consider the same system with $M$ users, and omit the noise term. The analytical performance of CSMA/CA is elaborated in \cite{bianchi2000performance}, which are omitted here due to the space limitation. Some main differences between the two protocols are listed as follow.

\begin{itemize}
\item {\bf Collision length.} In conventional CSMA/CA, the ``blindness'' in transmission results in long collision, which is typically a packet length. FD allows users to detect collision while transmitting. Thus, the average collision length $L_c$, as is derived in \eqref{lc}, is slightly more than one slot, which is sharply reduced compared with CSMA/CA.
\item {\bf Successful transmission length.} In CSMA, once a collision-free transmission begins, it can always be finished successfully without interruption. However, in FD-MAC, the transmission may get ceased due to false alarm, especially for long packets. According to~\eqref{ls}, if $L$ is sufficiently large, $L_s$ goes to $1/P_f$. Also, false alarm leads to unnecessary backoff and increase of contention window, which may further degrade the performance of FD-MAC. Thus, in FD-MAC, the design of an appropriate packet length should be carefully considered.
\end{itemize}

\vspace{-1em}
%%%%%%%%%%%%%%%%%%%%%%%%%%%%%%%%%%%%%%%%%%%%%%%
\section{Simulation Results}%
%%%%%%%%%%%%%%%%%%%%%%%%%%%%%%%%%%%%%%%%%%%%%%%

In this section, simulation results are presented to evaluate the performance of the FD-MAC protocol. We consider $M = 100$ users with $P_m=10^{-2}$ and $P_f=10^{-3}$, and the DIFS length is 2 slots. At the beginning of each simulation, we set the contention window of each user as the initial length (CW$_{\min}$), and run for $10^4$ transmission attempts to make sure that the system is fully developed. Then another $10^6$ transmission attempts are simulated to obtain the simulation results.

\begin{figure}[!t]
\centering
\includegraphics[width=3.3in]{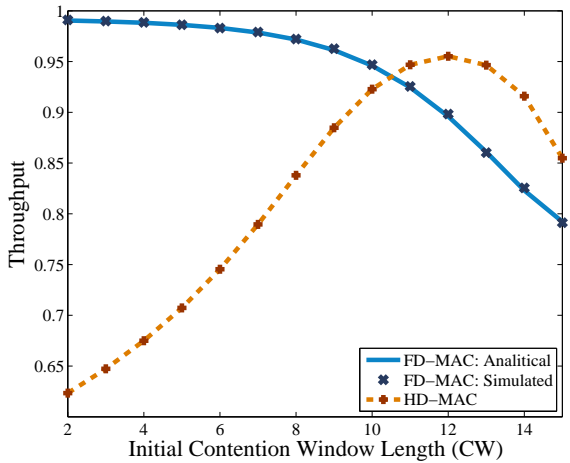}
\caption{Saturated throughput vs. initial contention window length CW$_{\min}$, where user number $M=100$, and CW$_{\max} = 2^{15}$ slots.} \label{cw}
\end{figure}

Fig.~\ref{cw} shows the saturated throughput for the proposed FD-MAC as well as the conventional CSMA/CA basic access scheme depending on different initial contention window sizes. We set the maximum contention window size as $2^{15}$ slots, and the packet length is fixed on 1000 slots. Both analytical~(the solid line) and simulated~(the crosses) performance of the FD-MAC are presented in Fig.~\ref{cw}, which match perfectly. It is shown that when the initial contention window length exceeds the number of users~(e.g., $2^7$ in this figure), the channel waste increases visibly, and throughput decreases due to the overlong backoff procedure. However, short initial CW length does not result in performance loss as the basic access scheme~(the dashed line). This is because the immediate backoff after collision in FD-MAC. Users can detect collision with high probabilities within 2 slots and backoff and adjust their contention window lengths accordingly instead of colliding for a whole packet unawarely. Also, Fig.~\ref{cw} shows that the channel usage can reach more than $99\%$ in FD-MAC, while in conventional protocol, the peak point cannot exceed $96\%$. It is seen that FD-MAC can achieve better performance.

\begin{figure}[!t]
\centering
\includegraphics[width=3.3in]{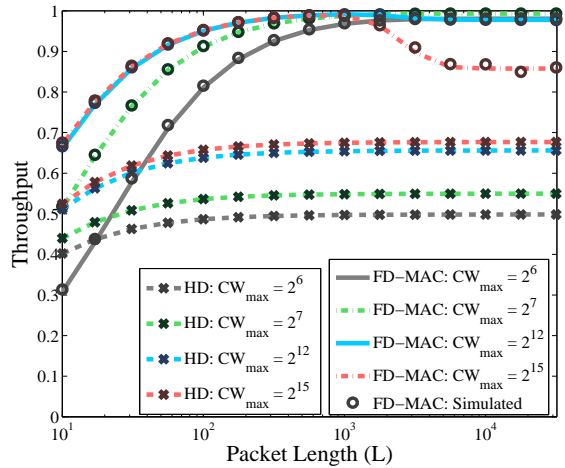}
\caption{Saturated throughput vs. packet length $L$, where CW$_{\min}=2^4$ slots, and $W_{\max}=\left\{2,3,8,11\right\}$} \label{l}
\end{figure}

In Fig.~\ref{l}, we consider the impact of packet length. We set minimum contention window length as CW$_{\min} = 2^4$ slots. The solid and dash-dotted lines with circles on them denote the performance of FD-MAC depending on different packet length, while the dashed lines with crosses represent the corresponding CSMA/CA scheme. It is shown that FD-MAC can achieve much higher throughput in most cases, and the asymptotic throughput can be close to 1. In the conventional scheme, the packet length does not affect sensing performance and backoff procedure. With the increase of packet length, the system throughput monotonously approaches the asymptotic performance, which, according to \cite{bianchi2000performance}, is the probability that a transmission is successful. On the other hand, in the proposed FD-MAC, when the packet is sufficiently long, a collision-free transmission is likely to stop in the middle due to false alarm and the user backoffs and prolongs the CW. When the system goes stable, it is likely that all users adopt the largest CW even if they rarely collide. Thus, when the packet length exceeds the average successful transmission length $L_s$, which is approximately $1/P_f = 10^3$ slots, the throughput depends mostly on CW$_{\max}$, with which it decreases. When CW$_{\max}$ is large enough~(e.g. $2^{12}$ or $2^{15}$ slots in Fig.~\ref{l}), the throughput may drop under long packet length, and the non-monotonicity can be observed.

\vspace{-1em}
%%%%%%%%%%%%%%%%%%%%%%%%%%%%%%%%%%%%%%%%%%%%%%%
\section{Conclusions}%
%%%%%%%%%%%%%%%%%%%%%%%%%%%%%%%%%%%%%%%%%%%%%%%

In this letter, we proposed a new FD-MAC protocol based on CSMA/CA basic access scheme for distributed access networks with FD transceivers. Basically, by the proposed protocol, users keep sensing and monitoring the channel usage state when they are transmitting. In the new protocol, the average collision length is largely reduced due to the continuous detection of channel usage and the high channel utilization ratio. Comparison between the new FD-MAC and conventional CSMA/CA shows the significant enhancement of throughput in both analytical and simulation results.

\vspace{-1em}
\bibliographystyle{IEEEtran}
\bibliography{IEEEabrv,REFS}

\begin{thebibliography}{20}

\bibitem{jain2011practical}
M.~Jain, J.~I.~Choi, T.~Kim, D.~Bharadia, S.~Seth, K.~Srinivasan, P.~Levis, S.~Katti, and P.~Sinha. ``Practical, Real-time, Full Duplex Wireless,'' in \emph{ACM MobiCom 2011}, New York, NY, Sep.~2011.

\bibitem{sahai2011pushing}
A.~Sahai, G.~Patel, and A.~Sabharwal, ``Pushing the limits of Full-duplex: Design and Real-time Implementation,'' Rice Univ., Houston, TX, USA, Rep. TREE1104, http://arxiv.org/pdf/1107.0607.pdf, Jul.~2011.

\bibitem{singh2011efficient}
N.~Singh, D.~Gunawardena, A.~Proutiere, B.~Radunovic, H.~V.~Balan, and P.~Key, ``Efficient and Fair MAC for Wireless Networks with Self-interference Cancellation,'' in \emph{Int. Symp. on Modeling and Optimization in Mobile, Ad Hoc and Wireless Networks~(WiOpt)}, pp.~94-101, Princeton, NJ, May~2011.

\bibitem{goyal2013distributed}
S.~Goyal, P.~Liu, O.~Gurbuz, E.~Erkip, and S.~Panwar, ``A Distributed MAC Protocol for Full Duplex Radio,'' in \emph{Asilomar Conference on Signals, Systems and Computers}, Pacific Grove, CA, Nov.~2013.

\bibitem{mypaper}
Y.~Liao, T.~Wang, L.~Song, and Z.~Han, ``Listen-and-Talk: Full-Duplex Cognitive Radio,'' in \emph{IEEE Globecom'2014}, Austin, TX, Dec.~2014.

\bibitem{bianchi2000performance}
G.~Bianchi,``Performance Analysis of the IEEE 802.11 Distributed Coordination Function,'' \emph{IEEE Trans. on Selected Areas in Comm.}, vol.~18, no.~3, pp.~535-547, Mar.~2000.


\end{thebibliography}

\end{document}